\documentclass[amsmath,amssymb,twocolumn]{revtex4-2}
\usepackage{graphicx}
\usepackage{subfig}
\usepackage{epsfig}
\usepackage{dcolumn}
\usepackage{multirow}
\usepackage{rotating}
\usepackage{verbatim}
\usepackage{bm}
\usepackage{color}
\usepackage{soul}
\usepackage{float}
\usepackage{hyperref}
\usepackage[utf8]{inputenc}

\def\lsim{\raise0.3ex\hbox{$<$\kern-0.75em\raise-1.1ex\hbox{$\sim$}}}
\def\gsim{\raise0.3ex\hbox{$>$\kern-0.75em\raise-1.1ex\hbox{$\sim$}}}

\def\beqa{\begin{eqnarray}}
\def\eeqa{\end{eqnarray}}

\begin{document}

\title{Universality of scaling entropy in charged hadron multiplicity distributions at the LHC}
\author{L.S. Moriggi$^{1}$}
\email[e-mail: ]{lucasmoriggi@unicentro.br}
\affiliation{$^{1}$ Universidade Estadual do Centro-Oeste (UNICENTRO), Campus Cedeteg, Guarapuava 85015-430, Brazil}

\author{F. S. Navarra} 
\email[e-mail: ]{ navarra@if.usp.br} 
\affiliation{Instituto de F\'{\i}sica, Universidade de S\~{a}o Paulo, 
  Rua do Mat\~ao, 1371, CEP 05508-090, Cidade Universit\'aria, 
  S\~{a}o Paulo, SP, Brazil.}

\author{M.V.T. Machado$^{2}$}
\email[e-mail: ]{magnus@if.ufrgs.br}
\affiliation{$^{2}$ High Energy Physics Phenomenology Group, GFPAE. Institute of Physics, Federal University of Rio Grande do Sul (UFRGS)\\
Caixa Postal 15051, CEP 91501-970, Porto Alegre, RS, Brazil} 

\begin{abstract}
In this work, we investigate the scaling behavior of the entropy associated with the charged hadron multiplicity distribution \( P(N) \) in proton–proton collisions at the LHC. We show that the growth of this entropic indicator as a function of the Bjorken-\( x \) variable exhibits universal behavior, consistent with observations from deep inelastic scattering (DIS). This universality suggests that the entropy scaling is a property of the initial state and reflects the diffusive nature of gluon dynamics at small \( x \). Furthermore, we demonstrate that high-multiplicity events are not accurately described by traditional KNO scaling and require a more precise description based on a diffusion scaling framework. This new scaling emerges naturally from the universal growth of partonic entropy and offers deeper insight into the dynamics of particle production in high-energy hadronic collisions.
\end{abstract}

\maketitle


\section{Introduction}

Understanding the probability distribution \( P(N) \) for producing \( N \) charged hadrons in high-energy collisions is fundamental to unraveling the dynamics of quarks and gluons during the initial and final stages of the collision process. Although \( P(N) \) is one of the most extensively studied observables in high-energy physics, recent LHC data have pushed the limits of our theoretical understanding. In particular, \emph{high-multiplicity events} — in which the number of produced particles greatly exceeds the mean — reveal significant discrepancies between theoretical predictions and experimental observations. These large fluctuations challenge conventional QCD expectations and expose tensions between standard event generators and data, not only in proton-proton (\(pp\)) but also in heavy-ion (\(AA\)) collisions.

The spectral shape of charged hadrons exhibits a strong dependence on event multiplicity, in a manner incompatible with existing QCD-inspired models and event generators~\cite{ALICE:2019dfi}. Specifically, the variation of transverse momentum (\(p_T\)) spectra across multiplicity classes leads to behaviors that are difficult to reconcile with current theoretical frameworks.

Without proper theoretical control of multiplicities in \(pp\) collisions, it becomes increasingly unrealistic to interpret final-state observables in heavy-ion collisions with confidence. In fact, phenomena traditionally attributed to quark-gluon plasma (QGP) formation may originate from a lack of understanding of the underlying gluon dynamics in the initial state. A recent ALICE study on multiplicities in \(pPb\) collisions~\cite{ALICE:2025woy} reveals large discrepancies between the data and the predictions from widely used Monte Carlo event generators. It also shows incompatibility with the GLASMA model \cite{Schenke:2013dpa}, highlighting serious limitations in current theoretical approaches. Moreover, several QGP-like signals observed in \(AA\) collisions are now being detected in small systems like \(pp\) and \(pPb\) collisions~\cite{ALICE:2025woy}, raising fundamental questions about their true origin. This convergence of phenomena across system sizes constitutes one of the most urgent open problems in high-energy hadron physics.

In the study of multiplicity distributions, one important subject of discussion is the famous 
Koba-Nielsen-Olesen (KNO) scaling \cite{Polyakov:1970lyy,Koba:1972ng}, which is expected to be valid at asymptotically high energies. 
From the experimental side, KNO scaling is simply the fact that the function $\langle N \rangle \, P(N)$ plotted against the variable $ N/\langle N\rangle$ is independent of collision energy $\sqrt{s}$. 
This prediction was made in the seventies (before QCD) and was based on general assumptions about high energy processes.  From the theoretical side, there were several attempts to understand it in terms of quark and gluon dynamics. This has proven to be a difficult task because particle production occurs in both soft (small momentum transfer) and hard (large momentum transfer) regimes, and we can only perform reliable calculations in the latter. 
Calculations with perturbative QCD (pQCD) have shown that KNO is found for particles produced in jets in 
high multiplicity events \cite{Dokshitzer:2025owq,Dokshitzer:2025fky}, where $N > \langle N \rangle$. This conclusion depends on the choice of resummation schemes 
and relies on some assumptions. For minimum bias events, which include a large contribution from soft events, 
KNO scaling was obtained in the framework of the Color Glass Condensate (CGC) approach \cite{Dumitru:2012yr,Dumitru:2012tw}
This behavior relied on the assumption that the effective theory describing color charge fluctuations at a
scale of the order of the saturation momentum is approximately Gaussian. Moreover, both non-linear saturation effects and running-coupling evolution were required to obtain KNO scaling. More recently, a multiplicity distribution  satisfying KNO scaling was derived by solving the Mueller dipole evolution equation in the double logarithm approximation \cite{Liu:2023eve}. In spite of all these efforts, a clear connection between KNO scaling
and QCD has not yet been established.

The multiplicity distributions measured by ALICE at different center-of-mass energies exhibit significant deviations from KNO scaling, particularly in the tails of the distribution where \( N/\langle N \rangle > 4 \)~\cite{ALICE:2015olq}. These deviations grow with increasing collision energy and are most pronounced in the high-multiplicity regime, where rare events dominate. The violation of KNO scaling reflects a breakdown of self-similarity and suggests the emergence of new dynamical scales not accounted for in early scaling hypotheses.

Various theoretical frameworks have been proposed to describe these deviations, including negative binomial distributions (NBDs) with soft and semi-hard components~\cite{Giovannini:1998zb} and multi-component models~\cite{Zborovsky:2018vyh}. While these approaches offer reasonable phenomenological fits, they typically involve a large number of free parameters that lack a clear interpretation in QCD and often exhibit inconsistent behavior when analyzed across different kinematic regimes.

The CMS Collaboration also investigated KNO scaling at different pseudorapidity windows~\cite{CMS:2010qvf}, showing stronger violations in broader rapidity intervals and at higher multiplicities. A comprehensive ATLAS study over a wide energy range (\( \sqrt{s} = 0.9 \) to 13~TeV)~\cite{ATLAS:2010jvh} confirmed that KNO violations become increasingly significant at lower energies and large multiplicities.

These findings raise several open questions:
\begin{itemize}
    \item Are the apparent collective effects in small systems the result of final-state interactions, or do they reflect intrinsic initial-state dynamics, such as gluon saturation?
    \item Can a universal scaling behavior be established across different collision systems, such as \(ep\), \(pp\), and \(AA\)?
    \item What mechanisms govern the rare, large-multiplicity events found in the tails of \( P(N) \)?
    \item Is KNO scaling a fundamental feature, or should it be replaced by a more general scaling law?
\end{itemize}

While a fully robust theoretical framework is still lacking, the idea of \emph{scaling laws} provides a promising bridge between empirical regularities in the data and theoretical models rooted in partonic dynamics. One such approach is based on \emph{small-$x$ scaling}, motivated by gluon saturation models. In this picture, the gluon distribution reaches a saturation scale \( Q_s \), which governs the typical transverse momentum and particle multiplicity. Evidence of such scaling behavior has been observed in inclusive particle production at both HERA~\cite{Golec-Biernat:1998zce,Stasto:2000er,Praszalowicz:2012zh,Gelis:2006bs,Moriggi:2020zbv} and the LHC~\cite{Moriggi:2020zbv,Moriggi:2024tbr,McLerran:2014apa,Osada:2019oor,Osada:2020zui,Praszalowicz:2013fsa}.

In~\cite{Moriggi:2024tiz,Moriggi:2024tbr}, a new observable called \emph{scaling entropy} was introduced, proposed as a consequence of the universal growth of gluon distributions at small Bjorken-\(x\), resulting in the functional form:
\[
S(x) = \lambda \log(1/x) + \text{const}.
\]
Remarkably, the entropy extracted from the charged hadron multiplicity distribution in deep inelastic scattering (DIS) agrees with this prediction, and the growth exponent \( \lambda \) has been measured directly from the data.

In this study, we investigate the implications of \emph{scaling entropy} in the context of multiplicity distributions at the LHC. We perform a phenomenological analysis of the probability distribution \( P(N) \) from ALICE, CMS, and ATLAS data. We demonstrate that, after subtracting the soft component of the spectrum, the entropy of hadron production in \(pp\) collisions exhibits scaling behavior consistent with that found in HERA \(ep\) data. This suggests the presence of universal initial-state dynamics governing particle production.

As a consequence of this behavior, we argue that \emph{KNO scaling} must be replaced with a more accurate \emph{diffusion scaling}, which naturally emerges from the universal growth of parton distributions and entropy in the initial state. This new perspective offers a path toward a deeper understanding of high-multiplicity events and their connection to fundamental QCD dynamics.

\section{Theoretical framework and main predictions}
\label{sec:model}

In the dipole picture of deep inelastic scattering (DIS)~\cite{Nikolaev:1990ja}, the internal structure of the proton is probed via the scattering of a virtual photon. This process can be factorized into the convolution of the photon wave function and a dipole-target scattering amplitude \( \mathcal{N}(x, r) \), resulting in the total cross section:
\begin{equation} \label{eq:gammap}
    \sigma_{\gamma^* p}(x, Q^2) = \sigma_0 \int d^2\mathbf{r} \, dz \, |\psi_{\gamma}(r,z)|^2 \, \mathcal{N}(x, r),
\end{equation}
where \( \mathbf{r} \) is the transverse size of the dipole, \( z \) is the longitudinal momentum fraction carried by the quark, and \( \psi_{\gamma}(r,z) \) is the photon light-cone wave function.

The scattering amplitude \( \mathcal{N}(x, r) \) must respect unitarity in position space, i.e., \( \mathcal{N}(x, r) \leq 1 \), which ensures the saturation of interaction probability at large \( r \). This leads to the emergence of a dynamical scale known as the \emph{saturation scale}, \( Q_s(x) \), which marks the transition between dilute and dense partonic systems and is a decreasing function of Bjorken-\( x \).

In momentum space \( k_T\), the unitarity constraint manifests as the growth of the variance of the transverse momentum distribution with decreasing \( x \). In a parton cascade picture, emissions follow an evolution where the characteristic time scale is governed by the uncertainty principle, \( t \sim 1/x \), leading to anomalous diffusion in transverse momentum space:
\begin{equation} \label{eq:einstein1}
    \langle k_T^2(x) \rangle \sim \left( \frac{1}{x} \right)^{\lambda},
\end{equation}
where \( \lambda \) is related to the \emph{hard Pomeron intercept}, characterizing the rate of growth of partonic activity and gluon density at small \( x \).

We can define the Fourier transform,  $\mathcal{P}(x, k_T)$, of the scattering amplitude, \( \mathcal{N}(x, r) \). It is normalized to unity and can be interpreted as a probability distribution containing all information about the interaction process at the partonic level. The form of $\mathcal{P}(x, k_T)$ can be derived from the maximum entropy principle using Lagrange multipliers to optimize the distribution in a canonical ensemble. The most general form of nonextensive statistical mechanics, the Tsallis entropy \cite{Tsallis:1987eu}, is given by:
\begin{equation}\label{eq:tsallis}
    S_q = \int d^2k_T \frac{1 - [\mathcal{P}(x, k_T)]^q}{q - 1},
\end{equation}
with constraints:
\begin{equation} \label{eq:ktm}
    \langle k_T^2 \rangle_q = \frac{\int d^2k_T k_T^2 [\mathcal{P}(k_T)]^q}{\int d^2k_T [\mathcal{P}(k_T)]^q} = \beta^{-1}, \quad \int d^2k_T \mathcal{P}(k_T) = 1.
\end{equation}
The Tsallis q-index represents the degree of non-extensivity of the distribution. 

This framework leads to a stationary state with a power-law distribution, as proposed in Refs. \cite{Moriggi:2020zbv, Moriggi:2024tbr}. The Lagrange parameter $\beta$ can be linked to the scaling hypothesis:
\begin{equation}\label{eq:einstein}
    \langle k_T^2(x) \rangle_q \sim \beta^{-1} (x_s/x)^{\lambda}.
\end{equation}

Since $1/x$ corresponds to the time scale probed by the soft gluons, the interaction with gluons of varying energies results in an anomalous diffusion-like process. This connects $\langle k_T^2(x) \rangle_q$ to a generalized form of Einstein's relation for diffusion. Under these conditions, the probability distribution can be expressed in a scaling form:
\begin{equation}\label{eq:scaling}
    \mathcal{P}(x, k_T^2) \sim \frac{1}{x^{-\lambda}} f(k_T^2 / x^{-\lambda}).
\end{equation}

The resulting partonic entropy in the Tsallis formalism is given by:
\begin{equation} \label{eq:entroQ}
    S_q^{\text{parton}}(x) = \frac{1}{q-1} - \left( \frac{2-q}{q-1} \right)^q (\pi Q_s^2(x))^{1-q}.
\end{equation}

This general form has been used to establish a relationship between the saturation scale and the overlap area in $pp$ collisions \cite{Moriggi:2024tbr}. Scaling exists regardless of whether $q = 1$ or $q \neq 1$; however, integrated $k_T$ data cannot distinguish differences associated with high-$k_T$ degrees of freedom. Such effects are observable in $p_T$-differential cross sections.

The entropy associated with the initial-state partonic distribution is defined by:
\begin{equation}
    S^{\text{parton}}(x) = -\int \mathcal{P}(x, k_T) \log[\mathcal{P}(x, k_T)] \, d^2k_T, 
\end{equation}
which is the Boltzmann-Gibbs (BG) form  in $q=1$ limit. Under the scaling form~\eqref{eq:scaling}, the entropy exhibits a universal logarithmic behavior:
\begin{equation} \label{eq:parton_entropy}
    S^{\text{parton}}(x) = C + \lambda \log\left( \frac{1}{x} \right),
\end{equation}
where the constant \( C \) encodes all the model-dependent details of \( \mathcal{P} \), and the \( \lambda \log(1/x) \) term reflects a universal scaling behavior associated with the growth of gluon densities. This scaling relation underpins the anomalous diffusion-like behavior of gluons and connects the entropy to partonic momentum distributions.

\vspace{0.5em}
\noindent
Experimentally, we can define a related observable from the multiplicity distribution of charged hadrons:
\begin{equation} \label{eq:entroexp}
    S^{\text{mult}} = -\sum_N P(N) \log P(N),
\end{equation}
where \( P(N) \) is the probability of producing \( N \) charged hadrons in a given event class.

In \( pp \) collisions, \( S^{\text{mult}} \) depends on the pseudorapidity interval \( \eta \) in which the measurement is performed and on the center-of-mass energy \( \sqrt{s} \) of the collision. The relevant Bjorken-\( x \) value for the partonic subprocess can be estimated from the minimal transverse momentum \( p_T \) and the pseudorapidity of the produced particles:
\begin{equation}
    x_{a,b} = \frac{p_T}{\sqrt{s}} e^{\pm \eta}.
\end{equation}
Since most partonic interactions occur at low \( x \), the effective value of \( x \) for the subprocess is approximately:
\begin{equation} \label{eq:xval}
    x \approx \frac{e^{-\eta}}{\sqrt{s}},
\end{equation}
assuming \( p_T \sim 1 \) GeV for soft-scale interactions.

Substituting this into Eq.~\eqref{eq:parton_entropy}, we can write a phenomenological scaling form for the experimentally measured entropy:
\begin{equation} \label{eq:entropy_scaling}
    S^{\text{mult}}(\sqrt{s}, \eta) = \lambda \log\left( \frac{1}{x} \right) + f(\eta),
\end{equation}
where the first term depends only on the initial-state gluon dynamics through \( x \), and the second term \( f(\eta) \) may include contributions from hadronization and final-state effects independent of \( \sqrt{s} \). This scaling hypothesis provides a quantitative approach to test the principle of \emph{local parton-hadron duality} \cite{Dokshitzer:1987nm}, which posits that final-state hadron distributions closely reflect initial partonic configurations.

In DIS, Eq.~\eqref{eq:parton_entropy} can be directly tested thanks to the hard scale \( Q^2 \). In contrast, LHC data are often collected with low \( p_T \) cuts, where soft physics contributions may obscure the scaling behavior. To mitigate this, we divide particle production into two components: a soft component that is dominant at low multiplicities (\( N < N_{\text{min}} \)) and a semi-hard component that is dominant at high multiplicities (\( N > N_{\text{min}} \)). Scaling behavior is expected only in the latter.

While there is no universal definition for \( N_{\text{min}} \), our analysis (to be shown in the next section) reveals that for \( N_{\text{min}} \sim 10 \), the extracted growth rate \( \lambda \) stabilizes and becomes consistent with that observed in HERA DIS data, i.e., \( \lambda_{\text{LHC}} = \lambda_{\text{H1}} \). This observation supports the universality of the entropy growth law across different collision systems.

To quantify fluctuations, we also consider the variance of the multiplicity distribution:
\begin{equation} \label{eq:var}
    \text{Var}(N) = \sum ( N - \langle N \rangle)^2 P(N).
\end{equation}

\vspace{0.5em}
\noindent
The central goal of this work is to confront the theoretical prediction of parton scaling entropy in Eq.~\eqref{eq:parton_entropy} with the experimentally measured multiplicity entropy in Eq.~\eqref{eq:entroexp}, and to investigate whether a universal scaling behavior underlies particle production in high-energy hadron collisions.

Before presenting the main results, it is timely to compare the scaling entropy to recent entropy-based approaches to small-$x$ physics. One of them is the  entanglement (von Neumann) entropy evaluated in the Color Glass Condensate (CGC) approach, which is  obtained by taking into account  soft gluons in the wavefunction of a fast-moving hadron. The calculation can be done in the field basis or in the number representation basis \cite{Duan:2020jkz}. The leading contribution in terms of saturation scaling has the following form \cite{Duan:2020jkz}:
\begin{eqnarray}
\label{CGCSEE}
S_{EE}^{\mathrm{CGC}}\propto \frac{1}{2}S_{\perp}\frac{C_F}{4\pi}\tilde{Q}_s^2,
\end{eqnarray}
where $\tilde{Q}_s^2(x)=(9/4)(x_0/x)^{\lambda_g}$ and $S_{\perp}$ are the gluon saturation scale and the transverse size of the hadron/nucleus, respectively. The entropy  $S_{EE}^{\mathrm{CGC}}\sim x^{\lambda_g}$ has a stronger growth on $x$ compared to the scaling entropy, $S^{\mathrm{parton}}\sim \ln (1/x)^{\lambda}$.

Another formalism  is the Wehrl entropy in QCD \cite{Hagiwara:2017uaz}, which is the semiclassical analog of the von Neumann entropy. It is obtained in terms of quantum phase space distributions, such as Wigner or Husimi distributions.  The QCD Wigner distribution, $W (x,\vec{k},\vec{b})$, is a generalization of the usual collinear parton distribution functions and depends on parton transverse momentum, $\vec{k}$,  impact parameter, $\vec{b}$, and longitudinal momentum fraction, $x$. In case the Wigner distribution is positive definite, then the Wehrl entropy associated with the gluons can be defined in the following way \cite{Hatta:2016dxp},
\begin{eqnarray}
S_W &= &  -\int d^2bd^2k\, xW_{g}(x,k,b)\ln \left[ xW_{g}(x,k,b) \right], 
\label{WehrlS}
\end{eqnarray}
where $xg(x)$ is the usual collinear gluon distribution, with $xg(x) =  \int d^2bd^2k \,xW_{g}(x,k,b)$. The Weiszacker-Williams (WW) gluon Wigner distribution \cite{Kovchegov:1998bi,Dominguez:2011wm} can be written in terms of the forward  $S$-matrix of a QCD color dipole of transverse size $\vec{r}$, transverse momentum $\vec{k}$ at impact parameter $\vec{b}$ in the adjoint representation, ${\cal{S}}_A$:
\begin{eqnarray}
xW_g(x,k,b) = \frac{C_F}{2\pi^4 \alpha_s}\int d^2\vec{r} \,\frac{e^{i\vec{r}\cdot \vec{k}}}{r^2}\left( 1-{\cal{S}}_A(x,\vec{r},\vec{b}) \right). \nonumber\\
\end{eqnarray}
The WW Wigner distribution can be analytically evaluated in the case of a Gaussian form for the S-matrix, ${\cal{S}}_A(x,r,b)=\exp [-\vec{r}^2\tilde{Q}_s^2(x,b)/4]$, where $\tilde{Q}_s^2(x,b) =(N_c/C_F)Q_s^2(x,b)$ is the impact parameter dependent  gluon saturation scale. Specifically, for the Gaussian $S$-matrix one obtains the following Wehrl entropy \cite{Ramos:2020kaj},
\begin{eqnarray}
\label{SWap}
S_W  \approx  \frac{2N_c S_{\perp}}{6\pi^2\alpha_s}\,Q_s^2(x),
\end{eqnarray}
where the parametric behavior is $S_W\propto S_{\perp}Q_s^2(x)$. This behavior is quite similar to the von Neumann entropy computed in the CGC formalism. 

The parton entanglement entropy proposed in Ref. \cite{Kharzeev:2017qzs} (Kharzeev-Levin) is constructed based on the color dipole cascade equation in (1+1) dimensions.  At small-$x$ the relation between the entanglement entropy and the gluon distribution, $xg(x)$, is simple:
\begin{eqnarray}
 \label{SEESAT}
  S_{EE}^{KL}= \ln [xg(x)]\sim \ln x^{-\Delta},
 \end{eqnarray}
where $\Delta=4\alpha_SN_c\ln(2)/\pi$ is related to the hard Pomeron (BFKL) intercept. The main point is the identification of the probabilities of micro-states $p_n$ with the probabilities of finding $n$ dipoles inside the hadron $P_n(Y)$. The entanglement entropy, $S=-\sum_n P_n\ln(P_n)$, takes a similar form in the (3+1)-dimensional case.  The seminal paper \cite{Kharzeev:2017qzs} gave rise to a series of phenomenological analyzes of this entropy notion in recent years \cite{Tu:2019ouv,Gotsman:2020bjc,H1:2020zpd,Kharzeev:2021yyf,Levin:2021sbe,Zhang:2021hra,Liu:2022ohy,Liu:2022hto,Hentschinski:2022rsa,Dosch:2023bxj,Hentschinski:2023izh,Levin:2023mwl,Caputa:2024xkp,Hentschinski:2024gaa,Hatta:2024lbw,Levin:2024wtl,Ouchen:2025tta,Kutak:2025syp,Ramos:2020kaj}.   This entropy is very close to ours, where the parameter $\lambda$ is related, basically, to the energy dependence of the saturation scale $Q_s$. 

Finally, we mention the QCD dynamical entropy \cite{Peschanski:2012cw}, $\Sigma^{Y_0\rightarrow Y}$. It provides a microscopic definition of entropy analogous to the Boltzmann approach, based on the QCD dynamics of the CGC, and can be understood as a sort of relative entropy (RE) or Kullback–Leibler divergence \cite{10.2996/kmj/1138844604,Floerchinger:2020ogh}.  The microscopical point of view  of this entropy is related to the balance between branching and recombination in the saturation regime of dense QCD states. In this regime, gluons organize themselves into cells of typical saturation scale size $R_s(Y) = 1/Q_s(Y)$, where $Q_s \sim e^{\lambda Y}$ is the momentum saturation. From this, the non-linear energy evolution can be evaluated as a compression $R_s(Y_0) \rightarrow R_s(Y) < R_s(Y_0)$ for $Y_0 < Y$. The entropy is evaluated through the transverse momentum probability distributions $P(k,Y)$, where $k$ is the gluon transverse momentum. These distributions are obtained from the  unintegrated gluon distributions (UGD), $\varphi(k,Y)$. In the simplest case of a Gaussian in $k$ UGD and the geometric scaling property, one obtains \cite{Ramos:2022gia,Ramos:2022gia}:
\begin{eqnarray}
\Sigma^{Y_0\rightarrow Y}(\Delta Y)=& = &  \int d^2k \, P(k,Y) \ln \left[\frac{P(k,Y)}{P(k,Y_0)}\right], \nonumber \\
&=&2\left(e^{\lambda \Delta Y}-1-\lambda\Delta Y\right),
\label{protongbwdynamicalentropy}
\end{eqnarray}
where $\Delta Y = Y-Y_0$. It is hard to compare this notion of entropy to ours since it is a relative entropy. However, the main input is similar: a transverse momentum distribution probability.

\section{Results and discussions}
\label{sec:results}

\begin{table*}[t]
\centering
\caption{Extracted $\lambda$ values from different LHC data sets at various pseudorapidity windows and center-of-mass energies.}
\begin{tabular}{|l|c|c|c|c|c|}
\hline
\textbf{Data Set} & $p_T$ threshold (GeV)& $\eta_{\text{max}}$ & $\sqrt{s}$ (GeV) & $\lambda \pm \Delta\lambda$ & $\chi^2/\text{dof}$ \\
\hline
1 - CMS  &0.1 & 0.5 & 900, 2360, 7000              & $0.32 \pm 0.02$ & 0.0003     \\
1 - CMS  &0.1  & 1.0 & 900, 2360, 7000              & $0.34 \pm 0.02$ & 0.0098    \\
1 - CMS  &0.1  & 1.5 & 900, 2360, 7000              & $0.33 \pm 0.02$ & 0.0179   \\
1 - CMS  &0.1  & 2.0 & 900, 2360, 7000              & $0.32 \pm 0.02$ & 0.0058    \\
1 - CMS  &0.1  & 2.5 & 900, 2360, 7000              & $0.32 \pm 0.02$ & 0.0003     \\
2 - ATLAS &0.1  & 2.5 & 2760, 900, 7000, 8000, 13000 & $0.32 \pm 0.01$ & 1.1045 \\
3 - ALICE &0.15  & 0.8 & 2760, 5020, 7000, 8000, 13000 & $0.32 \pm 0.01$ & 0.9907  \\
4 - ALICE &~0.05  & 0.5 & 900, 2760, 7000, 8000        & $0.29 \pm 0.03$ & 0.4514  \\
4 - ALICE &~0.05  & 1.0 & 900, 2760, 7000, 8000        & $0.30 \pm 0.02$ & 0.4980  \\
4 - ALICE &~0.05  & 1.5 & 900, 2760, 7000, 8000        & $0.29 \pm 0.01$ & 0.5657  \\
\hline
\end{tabular}
\label{tab:lambda_fits}
\end{table*}

\section{Results}

\begin{figure}[t]
\includegraphics[width=1.0\linewidth]{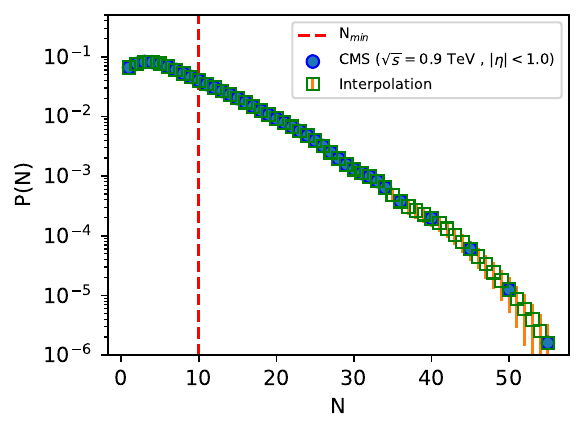}
\caption{Multiplicity data (full box) and interpolated values (unfilled) at some $N$ bins. The vertical line shows the cut imposed of $N_{min} = 10$.}\label{fig:PN}
\end{figure}

Different LHC collaborations adopt distinct methodologies for extracting the charged hadron multiplicity distribution \( P(N) \), often applying different kinematic cuts, pseudorapidity windows, and transverse momentum thresholds. For this analysis, we compiled data from CMS, ALICE, and ATLAS across several center-of-mass energies \( \sqrt{s} \) and pseudorapidity intervals \( \eta \), when available. We define four data sets for comparison:
\begin{itemize}
    \item \textbf{Set 1}: CMS data~\cite{CMS:2010qvf} with a large rapidity window, divided into five intervals from $\eta_{max}=0.5$ to $\eta_{max} = 2.5$. Non-single-diffractive (NSD) events were selected using a minimum-bias trigger and a lower $p_T$ threshold of $0.1$ GeV. Each rapidity window is provided for different values of $\sqrt{s}$.
    \item \textbf{Set 2}: ATLAS data~\cite{ATLAS:2010jvh,ATLAS:2016qux,ATLAS:2016zba} with a large rapidity window of $\eta_{max} = 2.5$ Events were collected using a single-arm minimum-bias trigger, with selections to reduce diffractive contributions. Only charged particles with transverse momentum \( p_T > 0.1 \) GeV were included.

    \item \textbf{Set 3}: ALICE data~\cite{ALICE:2022xip}, Inclusive primary charged particles with \( 0.15 \text{ GeV} < p_T < 10 \text{ GeV} \) were measured in a narrow pseudorapidity range of \( \eta_{max}=0.8 \). 
    \item \textbf{Set 4}: ALICE data~\cite{ALICE:2015olq} at extended rapidity coverage at different values of pseudorapidity from 0.5 to 1.5. Non single diffractive events with a lower \(p_T \) threshold of approximately 0.05 GeV.
\end{itemize}

Due to differences in analysis techniques and \( p_T \) thresholds, the multiplicity distributions in these sets are not directly comparable. A summary of the experimental characteristics and cuts applied is presented in Table~\ref{tab:lambda_fits}.

\begin{figure*}[t]
\includegraphics[width=1.0\linewidth]{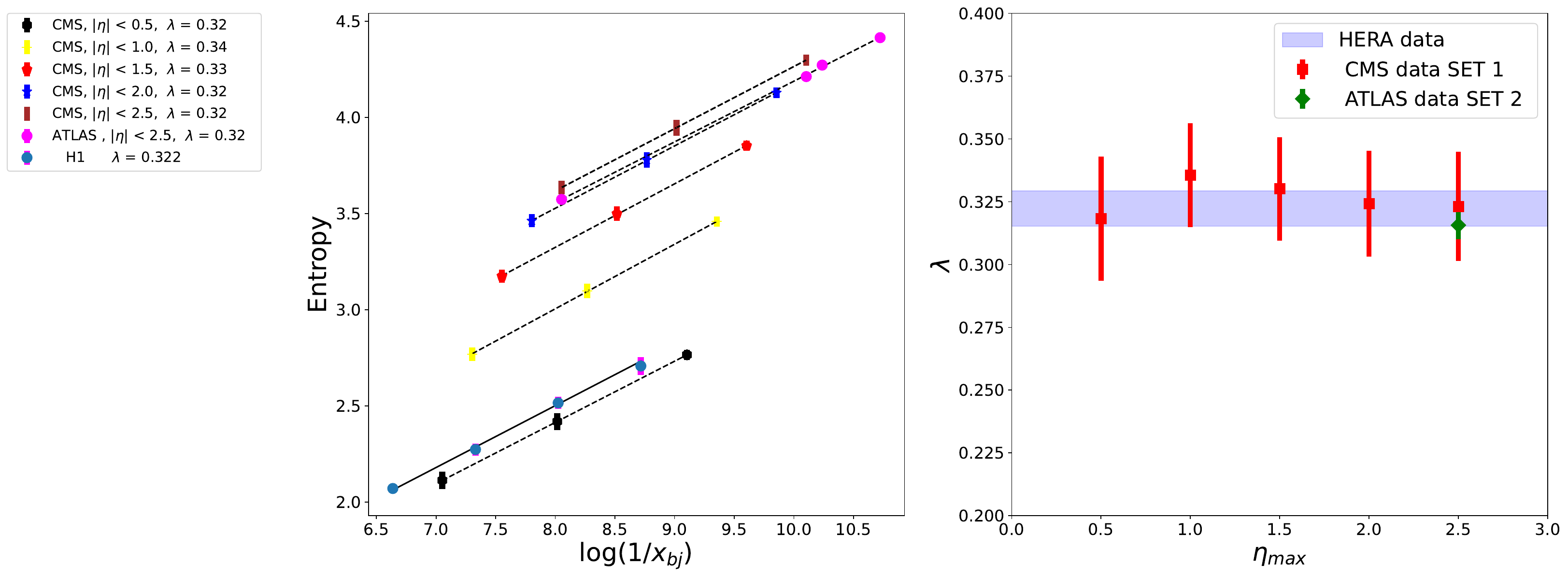}
\caption{Left side: Experimental entropy (color bars) compared with scaling entropy line as a function of $\log(1/x)$ in each rapidity bin when available. The full lines correspond to DIS entropy determined from H1 data. Right: resulting $\lambda$ compared with blue interval corresponding from H1 fit uncertainty.}\label{fig:entropy1}
\end{figure*}

\begin{figure*}[t]
\includegraphics[width=1.0\linewidth]{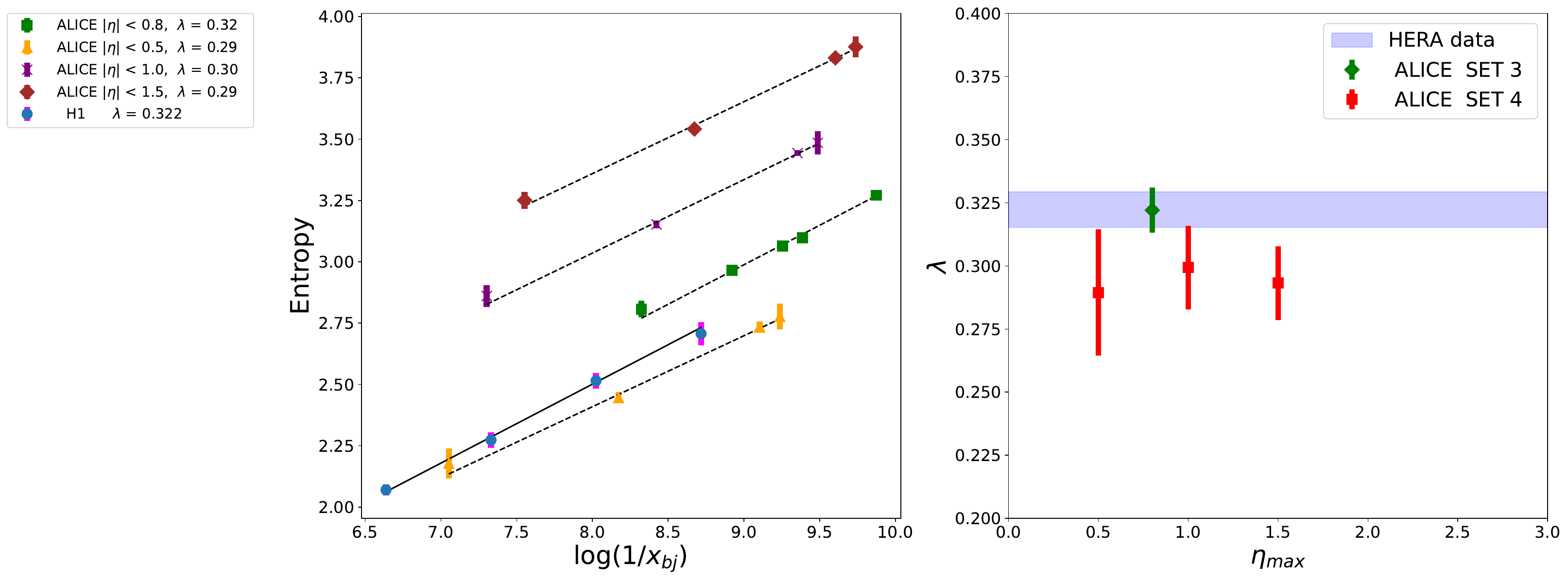}
\caption{The same as Fig. \ref{fig:entropy1} for the Sets 3 and 4.}\label{fig:entropy2}
\end{figure*}

In some cases (notably CMS), the \( P(N) \) values are not reported for all consecutive values of \( N \), which prevents direct computation of the entropy via Eq.~\eqref{eq:entroexp}. To address this, we perform a smooth interpolation using exponential cubic splines to estimate missing points and associated uncertainties. The procedure is illustrated in Fig.~\ref{fig:PN}, where the interpolated bins are shown unfilled, and a vertical dashed line indicates the cutoff value \( N_{\text{min}} \) applied to exclude soft contributions.

In ALICE Set 3 data, the available \( N \) range is limited. To ensure convergence in entropy calculations, we extrapolate the high-\( N \) tail of the distribution, which typically contributes about 5\% to the total entropy at \( \sqrt{s} = 0.9 \)~TeV but significantly impacts the extracted value of \( \lambda \). These preprocessing steps standardize the data sets for meaningful comparison.

We then fit the experimental entropy \( S^{\text{mult}} \) to the scaling form in Eq.~\eqref{eq:entropy_scaling}, assuming a logarithmic dependence on \( x \) for fixed \( \eta \). The fit extracts the entropy growth rate \( \lambda \) for each pseudorapidity interval. Results for Sets 1 and 2 are shown in Fig.~\ref{fig:entropy1}, where they are compared with the entropy extracted from deep inelastic scattering data \cite{Moriggi:2024tiz}.

We observe that the extracted values of \( \lambda \) across different LHC datasets and pseudorapidity windows are consistent, within uncertainties, with the DIS reference value \cite{Moriggi:2024tiz}:
\begin{equation} \label{eq:eresult}
    \lambda_{\text{DIS}} = 0.322 \pm 0.007.
\end{equation}

In particular, the CMS data at \( \sqrt{s} = 0.9 \)~TeV for the most central pseudorapidity range yield entropy values nearly identical to those in DIS. At larger rapidity windows, entropy increases with \( \sqrt{s} \), but the rate \( \partial S / \partial \log(1/x) \) remains consistent with the universal value \( \lambda \). This suggests that, despite experimental differences in \( P(N) \) extraction, the underlying entropy growth rate is a robust feature of QCD dynamics:
\begin{equation}
    \frac{\partial  S}{\partial\log(1/x)} = \lambda.
\end{equation}

Next, we analyze ALICE Sets 3 and 4 separately, as presented in Fig ~\ref{fig:entropy2}. Set 3 (central rapidity) shows good agreement with \( \lambda_{\text{DIS}} \), while Set 4 (extended rapidity) exhibits slightly smaller entropy growth rates, around 7\% lower than the other data sets. We believe that the difference is due to lower $p_T$ thresholds in Set 4, which allow for more soft events. Nevertheless, internal consistency is observed within the ALICE measurements at different \( \eta \).

An important complementary observable is the variance of the multiplicity distribution. For example, in the case of the negative binomial distribution (NBD), the variance is given by:
\begin{equation}
    \text{Var}(N) =  \langle N \rangle^2 + \frac{\langle N \rangle^2}{k},
\end{equation}
where \( k \) is the shape parameter, and both \( \langle N \rangle \) and \( k \) can vary with \( x \). Thus, NBD-based models imply that both variance and entropy encode information about partonic dynamics.

\begin{figure*}[t]
\includegraphics[width=0.7\linewidth]{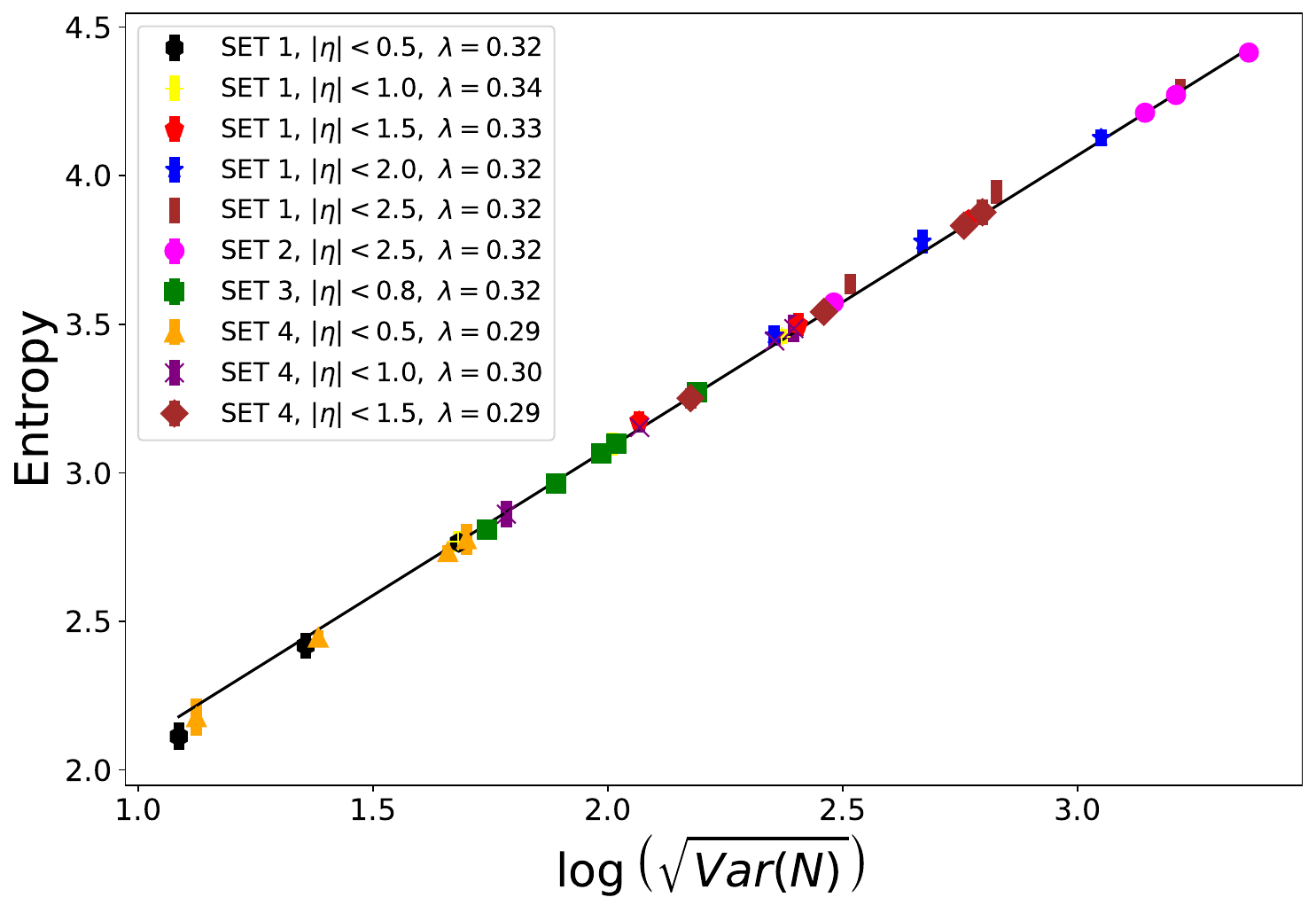}
\caption{Entropy as a function of the dispersion to each data set. The full line is a linear global fit.}\label{fig:variance}
\end{figure*}

Figure~\ref{fig:variance} shows the variation from the mean, quantified by the square root of the variance and entropy. We observe the linear behavior \( S = \log(\sqrt{Var(N)})  + cte \). 
This indicates that the entropy is governed not primarily by the average particle number, but by the event-by-event fluctuations around that mean. This dispersion originates from quantum fluctuations in the initial state, particularly in the small-$x$ regime. That means that the entropy effectively captures the spread of the parton distribution in momentum space.

\begin{figure*}[t]
\includegraphics[width=1.0\linewidth]{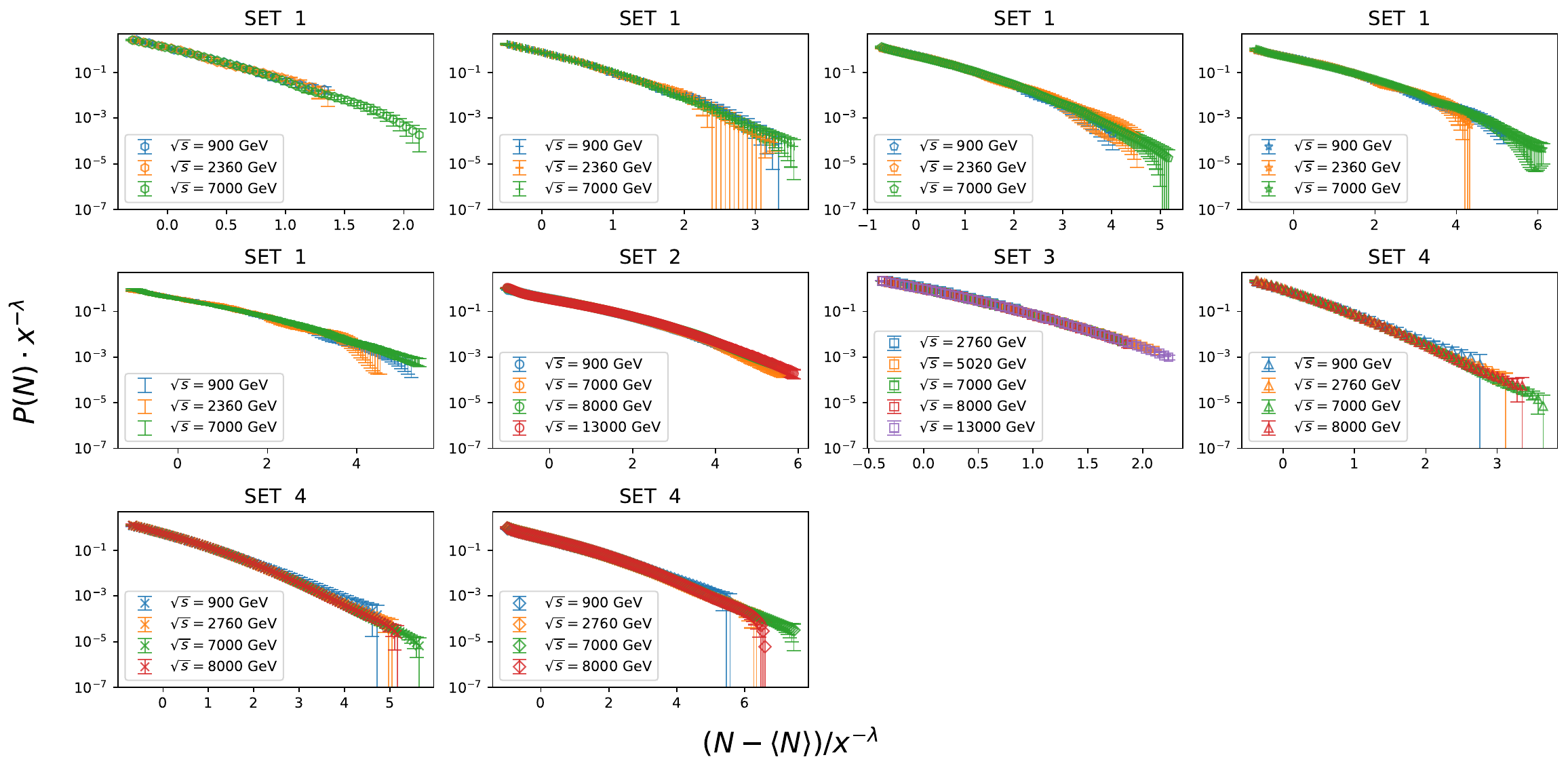}
\caption{Each data set scaling line at different rapidities.}\label{fig:diffusion}
\end{figure*}

Under the diffusion scaling hypothesis, the probability distribution for gluon \( k_T \) diffusion suggests a connection between entropy, variance, and initial-state \( x \) scaling. By analogy, the multiplicity distribution \( P(N) \) is expected to obey:
\begin{equation}
    P(N) \sim \frac{1}{x^{\lambda}} F\left( \frac{( N-\langle N \rangle ) }{x^{\lambda}} \right),
    \label{eq:equa14}
\end{equation}
where \( F \) is a universal scaling function. We do not formally derive a direct link between Eq. \ref{eq:scaling} and Eq. \ref{eq:equa14}; rather, this connection is proposed as a hypothesis to be tested against experimental data. Eq. \ref{eq:scaling} describes a random walk in transverse momentum space, driven by gluon activity in the initial state. In contrast, Eq. \ref{eq:equa14} suggests that this underlying partonic behavior manifests experimentally as fluctuations in the number of produced particles, following a similar scaling trend. In essence, the idea is that fluctuations in the transverse momentum of the initial state are mapped onto multiplicity fluctuations in the final state.

Figure~\ref{fig:diffusion} shows the diffusion scaling behavior for all data sets analyzed, confirming the remarkable universality of this scaling form across different experiments and kinematic regimes.

\begin{figure*}[t]
\includegraphics[width=1.0\linewidth]{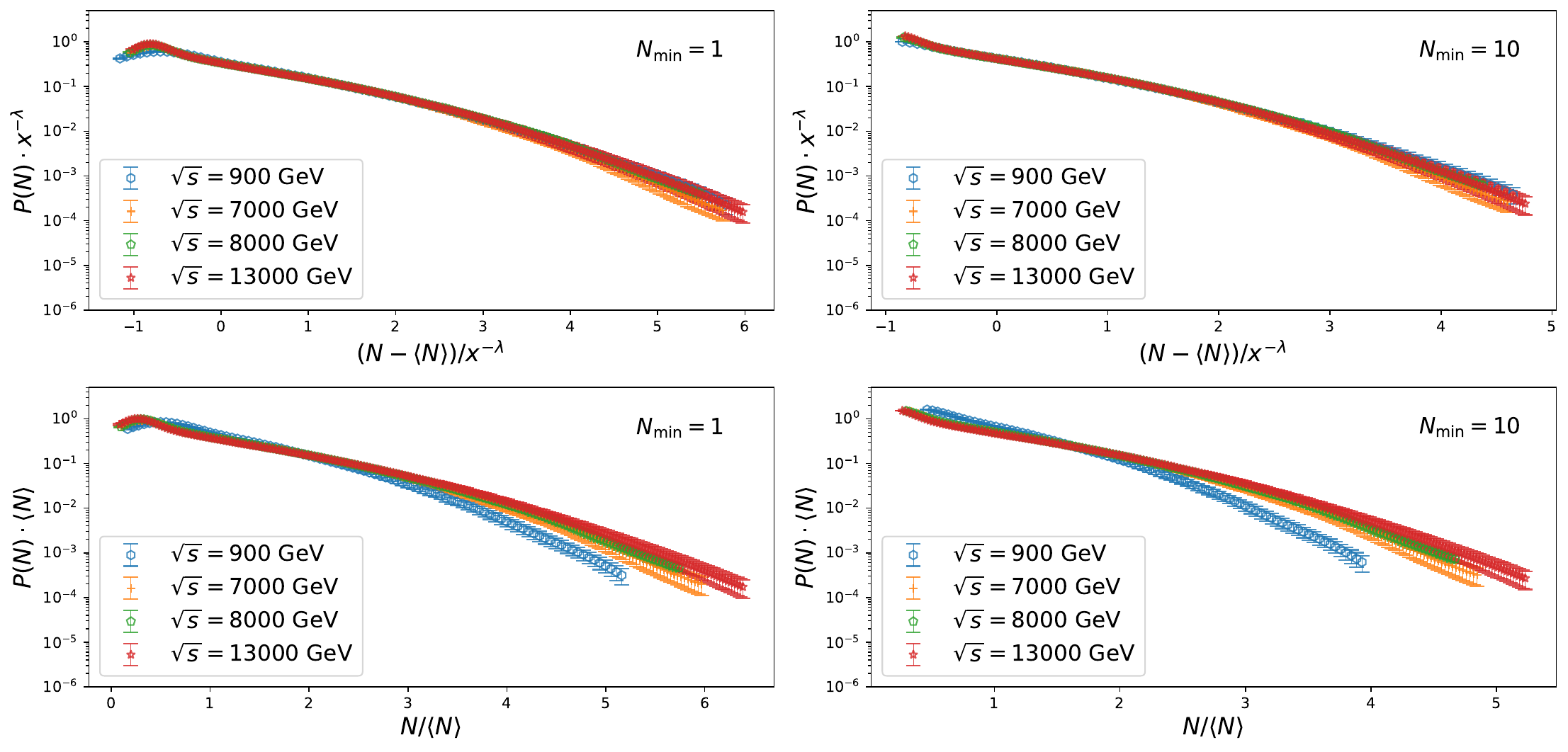}
\caption{Comparison between diffusion scaling (top) and KNO scaling (bottom) for two different values of \( N_{\min} \). Data from ATLAS Collaboration~\cite{ATLAS:2010jvh,ATLAS:2016qux,ATLAS:2016zba}, with $|\eta |< 2.5$. }\label{fig:scaling}
\end{figure*}

Additionally, we compare the traditional KNO scaling with the proposed diffusion scaling using ATLAS data~\cite{ATLAS:2010jvh,ATLAS:2016qux,ATLAS:2016zba}. Figure~\ref{fig:scaling} demonstrates that KNO scaling fails in the high-multiplicity tail (\( N \gg \langle N \rangle \)), while diffusion scaling remains valid. It is worth mentioning that the data description is qualitatively good; however, there are still clear discrepancies. The diffusive scaling curve in Fig. \ref{fig:diffusion} is obtained from the $\lambda$ values extracted from the entropy. Uncertainties in this measurement will also appear as discrepancies in the scaled probability $P(N)x^{-\lambda}$. The comparison is shown for multiple values of \( N_{\text{min}} \), and we find that the diffusive behavior becomes more evident when soft events are excluded. Notably, KNO scaling appears more reliable at lower multiplicities, suggesting that large fluctuations originate from partonic systems exhibiting diffusion-like behavior rather than self-similar scaling.

The violation of the KNO scaling is interesting. In the context of the Color Glass Condensate, the KNO scaling is related to the multiplicity distributions of soft gluons. It was shown in Ref. \cite{Dumitru:2012tw} that an approximately Gaussian effective theory (with running coupling evolution) describing color charge fluctuations at scales $\sim Q_s$ is able to reproduce the scaling. In particular, the NBD distribution can  be derived from this small-$x$ framework \cite{Dumitru:2012tw,Gelis:2009wh}.  The role played by the selection of the transverse momentum $p_T$ of produced charged particles within jets has been investigated recently in Ref. \cite{Germano:2024ier}.  Based on the ATLAS data, it has been shown that the low $p_T$ region does not exhibit KNO scaling, and the higher $p_T$ range gradually moves toward the scaling. On the other hand, all data can be described by a sub-Poissonian distribution, with a distinct $\delta$ parameter appearing in the birth rate $\mu_n$ in the cascade equations when the multiplicity is $n$, $\mu_n = \lambda(n+1)^{-\delta}+\sigma$. The different $\delta$ values from the fits suggest that the various transverse momentum  ranges are driven by heterogeneous dynamics.  Concerning high-$p_T$ jets, there is a general consensus that the  double logarithmic approximation (DLA) calculations in pQCD are able to describe the observed KNO scaling \cite{Liu:2023eve,Duan:2025ngi,Dokshitzer:2025owq}.

In an approach closest to ours, proposed by Levin and collaborators \cite{Levin:2024wtl,Levin:2023mwl,Gotsman:2020ubn,Gotsman:2020bjc}, the multiplicity distribution is related to the entanglement entropy of gluons in maximally entangled states. In the framework of the parton model, the high energy multiplicity distribution has the following form \cite{Kharzeev:2017qzs}: 
\begin{eqnarray}
 \label{eq:MDNB-Levin}
\frac{\sigma_n}{\sigma_{\rm in}}=\frac{\bar{N} }{\bar{N}\,+\,1}P^{\rm NBD}\left( 1, \bar{N},n\right) ,
\end{eqnarray}
where quantities $N$ (with $\bar{N}=N-1$), $\sigma_n$ and  $\sigma_{\rm in}$ are the average multiplicity, the cross section for producing $n$ hadrons, and the inelastic cross section, respectively.

The distribution of Eq. (\ref{eq:MDNB-Levin}) is able to correctly describe the data at multiplicities $n/\langle n \rangle \lesssim 4$ \cite{Gotsman:2020bjc}. On the other hand, it overestimates the data for large multiplicities. In order to mitigate this shortcoming, an approach that sums all Pomeron diagrams (based on $s$-channel unitarity of the QCD Reggeon Field Theory) was considered \cite{Kovner:2016iwy}. The new multiplicity distribution now takes the form:
\begin{eqnarray}
\label{eq:NEWPN-Levin}
P_n\,\,=\,\,\frac{1}{N}\left[ 1\,\,-\,\,\left( \frac{1}{N}\right)^{1 - \gamma}\right]^{n - 1},
\end{eqnarray}
where $\gamma$ is the dipole-dipole scattering amplitude in the Born approximation of pQCD.

 After this correction, Eq. (\ref{eq:NEWPN-Levin}), the system of partons with large multiplicities is suppressed in relation to the distribution in Eq. (\ref{eq:MDNB-Levin}), and LHC data are satisfactorily described \cite{Gotsman:2020bjc}. The key point is that the distribution of Eq. (\ref{eq:MDNB-Levin}) is obtained by considering the creation of a dilute system of partons at central rapidities, whereas the distribution in Eq. (\ref{eq:NEWPN-Levin}) is related to a dense system of partons.

\begin{figure}[t]
\includegraphics[width=1.0\linewidth]{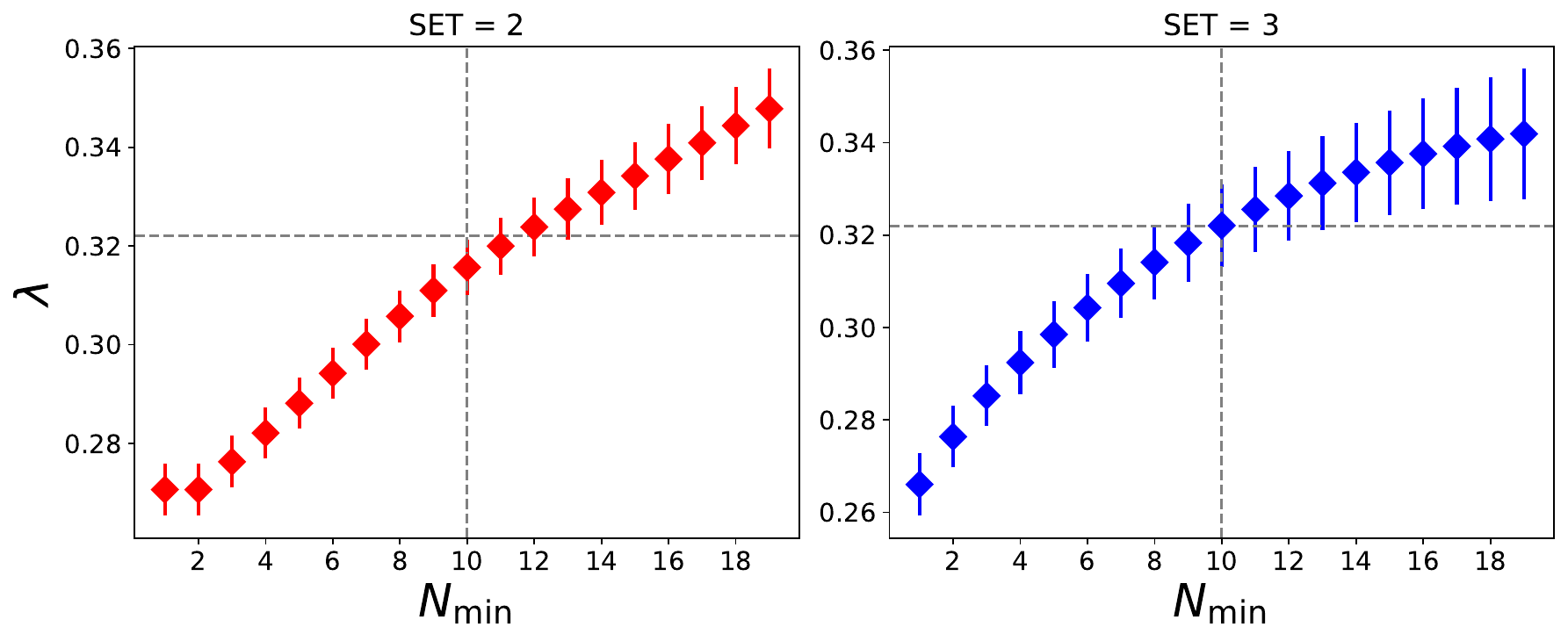}
\caption{The resulting $\lambda$ from SET 2 and SET 3 considering different cuts in multiplicity $N_{min}$.}\label{fig:Nmin}
\end{figure}

The final part of our analysis focuses on assessing the suitability of applying a fixed multiplicity cut at \( N_{\min} = 10 \). To evaluate the impact of this choice on the extracted values of \( \lambda \), we repeated the fitting procedure across a range of \( N_{\min} \) values.

When low-multiplicity events are included (i.e., small \( N_{\min} \)), the entropy growth rate is reduced, as expected from the dominance of soft processes and the slow increase of the total cross section in the absence of strong kinematic constraints. As we increase \( N_{\min} \), the contributions of semi-hard and hard processes become more significant, leading to faster entropy growth. This acceleration continues until a saturation point is reached around \( N_{\min} \sim 10 \), beyond which the growth rate stabilizes. This behavior is illustrated in Fig.~\ref{fig:Nmin}, indicating that above this threshold, particle production exhibits a universal scaling with respect to the Bjorken-\( x \) variable. This is also the point in which the values of $\lambda$ extracted from LHC data are equal to the HERA given by ~\eqref{eq:eresult}.

\section{SUMMARY AND CONCLUSIONS} \label{sec:conclusion}

This work was driven by the central questions mentioned in the Introduction section. Namely, whether a universal behavior governs charged hadron multiplicities across different high-energy collision systems—from HERA to the LHC—and whether the growth of the so-called scaling entropy exhibits universality rooted in initial-state QCD dynamics.

Our analysis demonstrates that, once low-multiplicity events are excluded, the entropy scaling observed in LHC data matches with remarkable precision the trends previously established in deep inelastic scattering at HERA. The extracted entropy growth rate, characterized by the parameter \( \lambda \), is consistent across experiments and pseudorapidity windows, reinforcing the idea of a common underlying mechanism.

Moreover, we showed that at large multiplicities—where fluctuations are most significant—the multiplicity distributions obey a precise diffusion scaling. This behavior strongly suggests that the observed growth in entropy and variance with energy is not a consequence of final-state effects or hadronization, but rather reflects a universal scaling pattern associated with partonic dynamics in the initial state.

In conclusion, our findings support the existence of a universal entropy growth law and diffusion scaling behavior in high-energy hadronic collisions, bridging observations from \( ep \) scattering at HERA to \( pp \) collisions at the LHC. This provides further evidence that the most extreme multiplicity events encode fundamental information about the small-\( x \) structure of hadrons and the dynamics of parton cascades.

Beyond its application to proton–proton collisions, the scaling entropy framework offers a valuable tool for disentangling initial- and final-state contributions in more complex systems, such as proton–nucleus and nucleus–nucleus collisions. Because entropy reflects early-stage partonic dynamics, it provides a clean observable largely insensitive to hadronization effects when analyzed above the soft-multiplicity threshold. This makes it a promising diagnostic for isolating initial-state physics in environments where final-state interactions, such as collective flow or quark-gluon plasma formation, are dominant.

Looking ahead, the scaling entropy method may also play an important role in future measurements at the upcoming Electron–Ion Collider (EIC), where precise control over the Bjorken-\( x \) and \( Q^2 \) kinematics will allow for stringent tests of universality and gluon saturation effects. Similarly, applying this framework to high-multiplicity events in heavy-ion collisions could help clarify which observables are rooted in initial-state structure versus those that emerge from final-state dynamics.

\vspace{1cm}

\section*{Acknowledgments}

MVTM acknowledges funding from the Brazilian agency Conselho Nacional de Desenvolvimento Científico e Tecnológico (CNPq) with the grant CNPq/303075/2022-8.  F.S.N. gratefully acknowledges the support from the Fundação de Amparo à Pesquisa  do Estado de São Paulo (FAPESP).

\bibliographystyle{h-physrev}
\bibliography{scalingentropy}

@article{Moriggi:2024tiz,
    author = "Moriggi, Lucas Soster and Machado, Magno Val\'erio Trindade",
    title = "{Precise determination of the Pomeron intercept via a scaling entropy analysis}",
    eprint = "2412.16348",
    archivePrefix = "arXiv",
    primaryClass = "hep-ph",
    doi = "10.1103/PhysRevD.111.014017",
    journal = "Phys. Rev. D",
    volume = "111",
    number = "1",
    pages = "014017",
    year = "2025"
}

@article{CMS:2010qvf,
    author = "Khachatryan, Vardan and others",
    collaboration = "CMS",
    title = "{Charged Particle Multiplicities in $pp$ Interactions at $\sqrt{s}=0.9$, 2.36, and 7 TeV}",
    eprint = "1011.5531",
    archivePrefix = "arXiv",
    primaryClass = "hep-ex",
    reportNumber = "CERN-PH-EP-2010-048, CMS-QCD-10-004",
    doi = "10.1007/JHEP01(2011)079",
    journal = "JHEP",
    volume = "01",
    pages = "079",
    year = "2011"
}

@article{ATLAS:2010jvh,
    author = "Aad, G. and others",
    collaboration = "ATLAS",
    title = "{Charged-particle multiplicities in pp interactions measured with the ATLAS detector at the LHC}",
    eprint = "1012.5104",
    archivePrefix = "arXiv",
    primaryClass = "hep-ex",
    reportNumber = "CERN-PH-EP-2010-079",
    doi = "10.1088/1367-2630/13/5/053033",
    journal = "New J. Phys.",
    volume = "13",
    pages = "053033",
    year = "2011"
}

@article{ATLAS:2016zba,
    author = "Aaboud, Morad and others",
    collaboration = "ATLAS",
    title = "{Charged-particle distributions at low transverse momentum in $\sqrt{s} = 13$  TeV $pp$ interactions measured with the ATLAS detector at the LHC}",
    eprint = "1606.01133",
    archivePrefix = "arXiv",
    primaryClass = "hep-ex",
    reportNumber = "CERN-EP-2016-099",
    doi = "10.1140/epjc/s10052-016-4335-y",
    journal = "Eur. Phys. J. C",
    volume = "76",
    number = "9",
    pages = "502",
    year = "2016"
}

@article{ATLAS:2016qux,
    author = "Aad, Georges and others",
    collaboration = "ATLAS",
    title = "{Charged-particle distributions in $pp$ interactions at $\sqrt{s}=$ 8 TeV measured with the ATLAS detector}",
    eprint = "1603.02439",
    archivePrefix = "arXiv",
    primaryClass = "hep-ex",
    reportNumber = "CERN-EP-2016-020",
    doi = "10.1140/epjc/s10052-016-4203-9",
    journal = "Eur. Phys. J. C",
    volume = "76",
    number = "7",
    pages = "403",
    year = "2016"
}

@article{ALICE:2022xip,
    author = "Acharya, Shreyasi and others",
    collaboration = "ALICE",
    title = "{Multiplicity dependence of charged-particle production in pp, p-Pb, Xe-Xe and Pb-Pb collisions at the LHC}",
    eprint = "2211.15326",
    archivePrefix = "arXiv",
    primaryClass = "nucl-ex",
    reportNumber = "CERN-EP-2022-266",
    doi = "10.1016/j.physletb.2023.138110",
    journal = "Phys. Lett. B",
    volume = "845",
    pages = "138110",
    year = "2023"
}

@article{ALICE:2015olq,
    author = "Adam, Jaroslav and others",
    collaboration = "ALICE",
    title = "{Charged-particle multiplicities in proton\textendash{}proton collisions at $\sqrt{s} = 0.9$ to 8 TeV}",
    eprint = "1509.07541",
    archivePrefix = "arXiv",
    primaryClass = "nucl-ex",
    reportNumber = "CERN-PH-EP-2015-259",
    doi = "10.1140/epjc/s10052-016-4571-1",
    journal = "Eur. Phys. J. C",
    volume = "77",
    number = "1",
    pages = "33",
    year = "2017"
}

@article{ALICE:2025woy,
    author = "Acharya, Shreyasi and others",
    collaboration = "ALICE",
    title = "{Charged-particle multiplicity distributions over a wide pseudorapidity range in p-Pb collisions at $\mathbf{\sqrt{s}_{NN} = 5.02}$ TeV}",
    eprint = "2502.18081",
    archivePrefix = "arXiv",
    primaryClass = "nucl-ex",
    reportNumber = "CERN-EP-2025-023",
    month = "2",
    year = "2025"
}

@article{ALICE:2019dfi,
    author = "Acharya, Shreyasi and others",
    collaboration = "ALICE",
    title = "{Charged-particle production as a function of multiplicity and transverse spherocity in pp collisions at $\sqrt{s} =5.02$ and 13 TeV}",
    eprint = "1905.07208",
    archivePrefix = "arXiv",
    primaryClass = "nucl-ex",
    reportNumber = "CERN-EP-2019-094",
    doi = "10.1140/epjc/s10052-019-7350-y",
    journal = "Eur. Phys. J. C",
    volume = "79",
    number = "10",
    pages = "857",
    year = "2019"
}

@article{Giovannini:1998zb,
    author = "Giovannini, Alberto and Ugoccioni, Roberto",
    title = "{Possible scenarios for soft and semihard components structure in central hadron hadron collisions in the TeV region}",
    eprint = "hep-ph/9810446",
    archivePrefix = "arXiv",
    reportNumber = "DFTT-36-98, FISIST-12-98-CENTRA",
    doi = "10.1103/PhysRevD.69.059903",
    journal = "Phys. Rev. D",
    volume = "59",
    pages = "094020",
    year = "1999",
    note = "[Erratum: Phys.Rev.D 69, 059903 (2004)]"
}

@article{Zborovsky:2018vyh,
    author = "Zborovsk\'y, I.",
    title = "{Three-component multiplicity distribution, oscillation of combinants and properties of clans in pp collisions at the LHC}",
    eprint = "1811.11230",
    archivePrefix = "arXiv",
    primaryClass = "hep-ph",
    doi = "10.1140/epjc/s10052-018-6287-x",
    journal = "Eur. Phys. J. C",
    volume = "78",
    number = "10",
    pages = "816",
    year = "2018"
}

@article{H1:2020zpd,
    author = "Andreev, V. and others",
    collaboration = "H1",
    title = "{Measurement of charged particle multiplicity distributions in DIS at HERA and its implication to entanglement entropy of partons}",
    eprint = "2011.01812",
    archivePrefix = "arXiv",
    primaryClass = "hep-ex",
    reportNumber = "DESY-20-176",
    doi = "10.1140/epjc/s10052-021-08896-1",
    journal = "Eur. Phys. J. C",
    volume = "81",
    number = "3",
    pages = "212",
    year = "2021"
}

@article{Dokshitzer:1987nm,
    author = "Dokshitzer, Yuri L. and Khoze, Valery A. and Troian, S. I. and Mueller, Alfred H.",
    title = "{QCD Coherence in High-Energy Reactions}",
    reportNumber = "CU-TP-374",
    doi = "10.1103/RevModPhys.60.373",
    journal = "Rev. Mod. Phys.",
    volume = "60",
    pages = "373",
    year = "1988"
}

@article{Praszalowicz:2012zh,
    author = "Praszalowicz, Michal and Stebel, Tomasz",
    title = "{Quantitative Study of Geometrical Scaling in Deep Inelastic Scattering at HERA}",
    eprint = "1211.5305",
    archivePrefix = "arXiv",
    primaryClass = "hep-ph",
    doi = "10.1007/JHEP03(2013)090",
    journal = "JHEP",
    volume = "03",
    pages = "090",
    year = "2013"
}

@article{Gelis:2006bs,
    author = "Gelis, F. and Peschanski, Robert B. and Soyez, G. and Schoeffel, L.",
    title = "{Systematics of geometric scaling}",
    eprint = "hep-ph/0610435",
    archivePrefix = "arXiv",
    doi = "10.1016/j.physletb.2007.01.055",
    journal = "Phys. Lett. B",
    volume = "647",
    pages = "376--379",
    year = "2007"
}

@article{McLerran:2014apa,
    author = "McLerran, Larry and Praszalowicz, Michal",
    title = "{Geometrical Scaling and the Dependence of the Average Transverse Momentum on the Multiplicity and Energy for the ALICE Experiment}",
    eprint = "1407.6687",
    archivePrefix = "arXiv",
    primaryClass = "hep-ph",
    doi = "10.1016/j.physletb.2014.12.046",
    journal = "Phys. Lett. B",
    volume = "741",
    pages = "246--251",
    year = "2015"
}

@article{Osada:2019oor,
    author = "Osada, Takeshi and Kumaoka, Takuya",
    title = "{Saturation momentum scale extracted from semi-inclusive transverse spectra in high-energy pp collisions}",
    eprint = "1904.10823",
    archivePrefix = "arXiv",
    primaryClass = "hep-ph",
    doi = "10.1103/PhysRevC.100.034906",
    journal = "Phys. Rev. C",
    volume = "100",
    number = "3",
    pages = "034906",
    year = "2019"
}

@article{Osada:2020zui,
    author = "Osada, Takeshi",
    title = "{Multiplicity-dependent saturation momentum in $p$-Pb collisions at 5.02 TeV}",
    eprint = "2011.00456",
    archivePrefix = "arXiv",
    primaryClass = "nucl-th",
    doi = "10.1103/PhysRevC.103.024911",
    journal = "Phys. Rev. C",
    volume = "103",
    number = "2",
    pages = "024911",
    year = "2021"
}

@article{Praszalowicz:2013fsa,
    author = "Praszalowicz, Michal",
    title = "{Geometrical scaling for identified particles}",
    eprint = "1308.5911",
    archivePrefix = "arXiv",
    primaryClass = "hep-ph",
    doi = "10.1016/j.physletb.2013.10.067",
    journal = "Phys. Lett. B",
    volume = "727",
    pages = "461--467",
    year = "2013"
}

@article{Golec-Biernat:1998zce,
    author = "Golec-Biernat, Krzysztof J. and Wusthoff, M.",
    title = "{Saturation effects in deep inelastic scattering at low Q**2 and its implications on diffraction}",
    eprint = "hep-ph/9807513",
    archivePrefix = "arXiv",
    reportNumber = "DTP-98-50",
    doi = "10.1103/PhysRevD.59.014017",
    journal = "Phys. Rev. D",
    volume = "59",
    pages = "014017",
    year = "1998"
}

@article{Stasto:2000er,
    author = "Stasto, A. M. and Golec-Biernat, Krzysztof J. and Kwiecinski, J.",
    title = "{Geometric scaling for the total gamma* p cross-section in the low x region}",
    eprint = "hep-ph/0007192",
    archivePrefix = "arXiv",
    reportNumber = "DESY-00-103",
    doi = "10.1103/PhysRevLett.86.596",
    journal = "Phys. Rev. Lett.",
    volume = "86",
    pages = "596--599",
    year = "2001"
}

@article{Moriggi:2020zbv,
    author = "Moriggi, L. S. and Peccini, G. M. and Machado, M. V. T.",
    title = "{Investigating the inclusive transverse spectra in high-energy $pp$ collisions in the context of geometric scaling framework}",
    eprint = "2005.07760",
    archivePrefix = "arXiv",
    primaryClass = "hep-ph",
    doi = "10.1103/PhysRevD.102.034016",
    journal = "Phys. Rev. D",
    volume = "102",
    number = "3",
    pages = "034016",
    year = "2020"
}

@article{Moriggi:2024tbr,
    author = "Moriggi, L. S. and Ramos, G. S. and Machado, M. V. T.",
    title = "{Multiplicity dependence of the pT-spectra for identified particles and its relationship with partonic entropy}",
    eprint = "2405.01712",
    archivePrefix = "arXiv",
    primaryClass = "hep-ph",
    doi = "10.1103/PhysRevD.110.034005",
    journal = "Phys. Rev. D",
    volume = "110",
    number = "3",
    pages = "034005",
    year = "2024"
}

@article{Koba:1972ng,
  author       = {Z. Koba and H. B. Nielsen and P. Olesen},
  title        = {Scaling of multiplicity distributions in high-energy hadron collisions},
  journal      = {Nucl. Phys. B},
  volume       = {40},
  pages        = {317--334},
  year         = {1972},
  doi          = {10.1016/0550-3213(72)90551-2},
  reportNumber = {NBI-HE-70-15},
}

@article{Nikolaev:1990ja,
  author       = {Nikolai N. Nikolaev and B.G. Zakharov},
  title        = {Colour transparency and scaling properties of nuclear shadowing in deep inelastic scattering},
  journal      = {Z. Phys. C},
  volume       = {49},
  pages        = {607--618},
  year         = {1991},
  doi          = {10.1007/BF01483577}
}

@article{Schenke:2013dpa,
    author = "Schenke, Bjoern and Tribedy, Prithwish and Venugopalan, Raju",
    title = "{Multiplicity distributions in p+p, p+A and A+A collisions from Yang-Mills dynamics}",
    eprint = "1311.3636",
    archivePrefix = "arXiv",
    primaryClass = "hep-ph",
    doi = "10.1103/PhysRevC.89.024901",
    journal = "Phys. Rev. C",
    volume = "89",
    number = "2",
    pages = "024901",
    year = "2014"
}

@article{Dumitru:2012tw,
    author = "Dumitru, Adrian and Petreska, Elena",
    title = "{KNO scaling from a nearly Gaussian action for small-x gluons}",
    eprint = "1209.4105",
    archivePrefix = "arXiv",
    primaryClass = "hep-ph",
    month = "9",
    year = "2012"
}

@article{Gelis:2009wh,
    author = "Gelis, F. and Lappi, T. and McLerran, L.",
    title = "{Glittering Glasmas}",
    eprint = "0905.3234",
    archivePrefix = "arXiv",
    primaryClass = "hep-ph",
    doi = "10.1016/j.nuclphysa.2009.07.004",
    journal = "Nucl. Phys. A",
    volume = "828",
    pages = "149--160",
    year = "2009"
}

@article{Germano:2024ier,
    author = "Germano, G. R. and Navarra, F. S. and Wilk, G. and Wlodarczyk, Z.",
    title = "{Emergence of Koba-Nielsen-Olsen scaling in multiplicity distributions in jets produced at the LHC}",
    eprint = "2406.04856",
    archivePrefix = "arXiv",
    primaryClass = "hep-ph",
    doi = "10.1103/PhysRevD.110.034026",
    journal = "Phys. Rev. D",
    volume = "110",
    number = "3",
    pages = "034026",
    year = "2024"
}

@article{Kharzeev:2017qzs,
    author = "Kharzeev, Dmitri E. and Levin, Eugene M.",
    title = "{Deep inelastic scattering as a probe of entanglement}",
    eprint = "1702.03489",
    archivePrefix = "arXiv",
    primaryClass = "hep-ph",
    doi = "10.1103/PhysRevD.95.114008",
    journal = "Phys. Rev. D",
    volume = "95",
    number = "11",
    pages = "114008",
    year = "2017"
}

@article{Levin:2024wtl,
    author = "Levin, Eugene",
    title = "{Particle production in a toy model: Multiplicity distribution and entropy}",
    eprint = "2412.02504",
    archivePrefix = "arXiv",
    primaryClass = "hep-ph",
    doi = "10.1103/PhysRevD.111.016019",
    journal = "Phys. Rev. D",
    volume = "111",
    number = "1",
    pages = "016019",
    year = "2025"
}

@article{Levin:2023mwl,
    author = "Levin, Eugene",
    title = "{Multiplicity distribution and entropy of produced gluons in deep inelastic scattering at high energies}",
    eprint = "2306.12055",
    archivePrefix = "arXiv",
    primaryClass = "hep-ph",
    doi = "10.1140/epjc/s10052-024-13008-w",
    journal = "Eur. Phys. J. C",
    volume = "84",
    number = "7",
    pages = "662",
    year = "2024"
}

@article{Gotsman:2020ubn,
    author = "Gotsman, E. and Levin, E.",
    title = "{High energy QCD: multiplicity dependence of quarkonia production}",
    eprint = "2008.10911",
    archivePrefix = "arXiv",
    primaryClass = "hep-ph",
    doi = "10.1140/epjc/s10052-020-08775-1",
    journal = "Eur. Phys. J. C",
    volume = "81",
    number = "2",
    pages = "99",
    year = "2021"
}

@article{Gotsman:2020bjc,
    author = "Gotsman, E. and Levin, E.",
    title = "{High energy QCD: multiplicity distribution and entanglement entropy}",
    eprint = "2006.11793",
    archivePrefix = "arXiv",
    primaryClass = "hep-ph",
    doi = "10.1103/PhysRevD.102.074008",
    journal = "Phys. Rev. D",
    volume = "102",
    number = "7",
    pages = "074008",
    year = "2020"
}

@article{Kovner:2016iwy,
    author = "Kovner, Alex and Levin, Eugene and Lublinsky, Michael",
    title = "{QCD unitarity constraints on Reggeon Field Theory}",
    eprint = "1605.03251",
    archivePrefix = "arXiv",
    primaryClass = "hep-ph",
    doi = "10.1007/JHEP08(2016)031",
    journal = "JHEP",
    volume = "08",
    pages = "031",
    year = "2016"
}

@article{Liu:2023eve,
    author = "Liu, Yizhuang and Nowak, Maciej A. and Zahed, Ismail",
    title = "{Universality of Koba-Nielsen-Olesen scaling in QCD at high energy and entanglement}",
    eprint = "2302.01380",
    archivePrefix = "arXiv",
    primaryClass = "hep-ph",
    month = "2",
    year = "2023"
}

@article{Duan:2025ngi,
    author = "Duan, Xiang-Pan and Chen, Lin and Ma, Guo-Liang and Salgado, Carlos A. and Wu, Bin",
    title = "{KNO scaling in quark and gluon jets at the LHC}",
    eprint = "2503.24200",
    archivePrefix = "arXiv",
    primaryClass = "hep-ph",
    month = "3",
    year = "2025"
}

@article{Dokshitzer:2025owq,
    author = "Dokshitzer, Yu. L. and Webber, B. R.",
    title = "{Hadron multiplicity fluctuations in perturbative QCD}",
    eprint = "2505.00652",
    archivePrefix = "arXiv",
    primaryClass = "hep-ph",
    month = "5",
    year = "2025"
}

@article{Dokshitzer:2025fky,
    author = "Dokshitzer, Yu. L. and Webber, B. R.",
    title = "{QCD-inspired description of multiplicity distributions in jets}",
    eprint = "2507.07691",
    archivePrefix = "arXiv",
    primaryClass = "hep-ph",
    month = "7",
    year = "2025"
}

@article{Dumitru:2012yr,
    author = "Dumitru, Adrian and Nara, Yasushi",
    title = "{KNO scaling of fluctuations in pp and pA, and eccentricities in heavy-ion collisions}",
    eprint = "1201.6382",
    archivePrefix = "arXiv",
    primaryClass = "nucl-th",
    reportNumber = "RBRC-941",
    doi = "10.1103/PhysRevC.85.034907",
    journal = "Phys. Rev. C",
    volume = "85",
    pages = "034907",
    year = "2012"
}

@article{Duan:2020jkz,
    author = "Duan, Haowu and Akkaya, Candost and Kovner, Alex and Skokov, Vladimir V.",
    title = "{Entanglement, partial set of measurements, and diagonality of the density matrix in the parton model}",
    eprint = "2001.01726",
    archivePrefix = "arXiv",
    primaryClass = "hep-ph",
    doi = "10.1103/PhysRevD.101.036017",
    journal = "Phys. Rev. D",
    volume = "101",
    number = "3",
    pages = "036017",
    year = "2020"
}

@article{Hagiwara:2017uaz,
    author = "Hagiwara, Yoshikazu and Hatta, Yoshitaka and Xiao, Bo-Wen and Yuan, Feng",
    title = "{Classical and quantum entropy of parton distributions}",
    eprint = "1801.00087",
    archivePrefix = "arXiv",
    primaryClass = "hep-ph",
    doi = "10.1103/PhysRevD.97.094029",
    journal = "Phys. Rev. D",
    volume = "97",
    number = "9",
    pages = "094029",
    year = "2018"
}

@article{Hatta:2016dxp,
    author = "Hatta, Yoshitaka and Xiao, Bo-Wen and Yuan, Feng",
    title = "{Probing the Small- x Gluon Tomography in Correlated Hard Diffractive Dijet Production in Deep Inelastic Scattering}",
    eprint = "1601.01585",
    archivePrefix = "arXiv",
    primaryClass = "hep-ph",
    reportNumber = "YITP-16-1",
    doi = "10.1103/PhysRevLett.116.202301",
    journal = "Phys. Rev. Lett.",
    volume = "116",
    number = "20",
    pages = "202301",
    year = "2016"
}

@article{Kovchegov:1998bi,
    author = "Kovchegov, Yuri V. and Mueller, Alfred H.",
    title = "{Gluon production in current nucleus and nucleon - nucleus collisions in a quasiclassical approximation}",
    eprint = "hep-ph/9802440",
    archivePrefix = "arXiv",
    reportNumber = "CU-TP-876",
    doi = "10.1016/S0550-3213(98)00384-8",
    journal = "Nucl. Phys. B",
    volume = "529",
    pages = "451--479",
    year = "1998"
}

@article{Dominguez:2011wm,
    author = "Dominguez, Fabio and Marquet, Cyrille and Xiao, Bo-Wen and Yuan, Feng",
    title = "{Universality of Unintegrated Gluon Distributions at small x}",
    eprint = "1101.0715",
    archivePrefix = "arXiv",
    primaryClass = "hep-ph",
    doi = "10.1103/PhysRevD.83.105005",
    journal = "Phys. Rev. D",
    volume = "83",
    pages = "105005",
    year = "2011"
}

@article{Ramos:2020kaj,
    author = "Ramos, G. S. and Machado, M. V. T.",
    title = "{Investigating entanglement entropy at small-$x$ in DIS off protons and nuclei}",
    eprint = "2003.05008",
    archivePrefix = "arXiv",
    primaryClass = "hep-ph",
    doi = "10.1103/PhysRevD.101.074040",
    journal = "Phys. Rev. D",
    volume = "101",
    number = "7",
    pages = "074040",
    year = "2020"
}

@article{Ramos:2022gia,
    author = "Ramos, G. S. and Machado, M. V. T.",
    title = "{Investigating the QCD dynamical entropy in high-energy hadronic collisions}",
    eprint = "2203.10986",
    archivePrefix = "arXiv",
    primaryClass = "hep-ph",
    doi = "10.1103/PhysRevD.105.094009",
    journal = "Phys. Rev. D",
    volume = "105",
    number = "9",
    pages = "094009",
    year = "2022"
}

@article{Peschanski:2012cw,
    author = "Peschanski, Robi",
    title = "{Dynamical entropy of dense QCD states}",
    eprint = "1211.6911",
    archivePrefix = "arXiv",
    primaryClass = "hep-ph",
    doi = "10.1103/PhysRevD.87.034042",
    journal = "Phys. Rev. D",
    volume = "87",
    number = "3",
    pages = "034042",
    year = "2013"
}

@article{10.2996/kmj/1138844604,
author = {Hisaharu Umegaki},
title = {{Conditional expectation in an operator algebra. IV. Entropy and information}},
volume = {14},
journal = {Kodai Mathematical Seminar Reports},
number = {2},
publisher = {Institute of Science Tokyo, Department of Mathematics},
pages = {59 -- 85},
year = {1962},
doi = {10.2996/kmj/1138844604},
URL = {https://doi.org/10.2996/kmj/1138844604}
}

@article{Floerchinger:2020ogh,
    author = "Floerchinger, Stefan and Haas, Tobias",
    title = "{Thermodynamics from relative entropy}",
    eprint = "2004.13533",
    archivePrefix = "arXiv",
    primaryClass = "cond-mat.stat-mech",
    doi = "10.1103/PhysRevE.102.052117",
    journal = "Phys. Rev. E",
    volume = "102",
    number = "5",
    pages = "052117",
    year = "2020"
}

@article{Polyakov:1970lyy,
    author = "Polyakov, A. M.",
    title = "{A Similarity Hypothesis in the Strong Interactions. I. Multiple Hadron Production in e+e- Annihilation}",
    journal = "Sov. Phys. JETP",
    volume = "32",
    pages = "296--301",
    year = "1971"
}

@article{Tsallis:1987eu,
    author = "Tsallis, Constantino",
    title = "{Possible Generalization of Boltzmann-Gibbs Statistics}",
    reportNumber = "CBPF-NF-062-87",
    doi = "10.1007/BF01016429",
    journal = "J. Statist. Phys.",
    volume = "52",
    pages = "479--487",
    year = "1988"
}

@article{Tu:2019ouv,
    author = "Tu, Zhoudunming and Kharzeev, Dmitri E. and Ullrich, Thomas",
    title = "{Einstein-Podolsky-Rosen Paradox and Quantum Entanglement at Subnucleonic Scales}",
    eprint = "1904.11974",
    archivePrefix = "arXiv",
    primaryClass = "hep-ph",
    doi = "10.1103/PhysRevLett.124.062001",
    journal = "Phys. Rev. Lett.",
    volume = "124",
    number = "6",
    pages = "062001",
    year = "2020"
}

@article{Kharzeev:2021yyf,
    author = "Kharzeev, Dmitri E. and Levin, Eugene",
    title = "{Deep inelastic scattering as a probe of entanglement: Confronting experimental data}",
    eprint = "2102.09773",
    archivePrefix = "arXiv",
    primaryClass = "hep-ph",
    doi = "10.1103/PhysRevD.104.L031503",
    journal = "Phys. Rev. D",
    volume = "104",
    number = "3",
    pages = "L031503",
    year = "2021"
}

@article{Levin:2021sbe,
    author = "Levin, Eugene",
    title = "{Multiplicity distribution of dipoles in QCD from the Le-Mueller-Munier equation}",
    eprint = "2106.06967",
    archivePrefix = "arXiv",
    primaryClass = "hep-ph",
    doi = "10.1103/PhysRevD.104.056025",
    journal = "Phys. Rev. D",
    volume = "104",
    number = "5",
    pages = "056025",
    year = "2021"
}

@article{Zhang:2021hra,
    author = "Zhang, Kun and Hao, Kun and Kharzeev, Dmitri and Korepin, Vladimir",
    title = "{Entanglement entropy production in deep inelastic scattering}",
    eprint = "2110.04881",
    archivePrefix = "arXiv",
    primaryClass = "quant-ph",
    doi = "10.1103/PhysRevD.105.014002",
    journal = "Phys. Rev. D",
    volume = "105",
    number = "1",
    pages = "014002",
    year = "2022"
}

@article{Liu:2022ohy,
    author = "Liu, Yizhuang and Nowak, Maciej A. and Zahed, Ismail",
    title = "{Entanglement entropy and flow in two-dimensional QCD: Parton and string duality}",
    eprint = "2202.02612",
    archivePrefix = "arXiv",
    primaryClass = "hep-ph",
    doi = "10.1103/PhysRevD.105.114027",
    journal = "Phys. Rev. D",
    volume = "105",
    number = "11",
    pages = "114027",
    year = "2022"
}

@article{Liu:2022hto,
    author = "Liu, Yizhuang and Nowak, Maciej A. and Zahed, Ismail",
    title = "{Rapidity evolution of the entanglement entropy in quarkonium: Parton and string duality}",
    eprint = "2203.00739",
    archivePrefix = "arXiv",
    primaryClass = "hep-ph",
    doi = "10.1103/PhysRevD.105.114028",
    journal = "Phys. Rev. D",
    volume = "105",
    number = "11",
    pages = "114028",
    year = "2022"
}

@article{Hentschinski:2022rsa,
    author = "Hentschinski, Martin and Kutak, Krzysztof and Straka, Robert",
    title = "{Maximally entangled proton and charged hadron multiplicity in Deep Inelastic Scattering}",
    eprint = "2207.09430",
    archivePrefix = "arXiv",
    primaryClass = "hep-ph",
    reportNumber = "IFJPAN-IV-2022-22",
    doi = "10.1140/epjc/s10052-022-11122-1",
    journal = "Eur. Phys. J. C",
    volume = "82",
    number = "12",
    pages = "1147",
    year = "2022"
}

@article{Dosch:2023bxj,
    author = "Dosch, Hans Gunter and de Teramond, Guy F. and Brodsky, Stanley J.",
    title = "{Entropy from entangled parton states and high-energy scattering behavior}",
    eprint = "2304.14207",
    archivePrefix = "arXiv",
    primaryClass = "hep-ph",
    reportNumber = "SLAC-PUB-17727",
    doi = "10.1016/j.physletb.2024.138521",
    journal = "Phys. Lett. B",
    volume = "850",
    pages = "138521",
    year = "2024"
}

@article{Hentschinski:2023izh,
    author = "Hentschinski, Martin and Kharzeev, Dmitri E. and Kutak, Krzysztof and Tu, Zhoudunming",
    title = "{Probing the Onset of Maximal Entanglement inside the Proton in Diffractive Deep Inelastic Scattering}",
    eprint = "2305.03069",
    archivePrefix = "arXiv",
    primaryClass = "hep-ph",
    doi = "10.1103/PhysRevLett.131.241901",
    journal = "Phys. Rev. Lett.",
    volume = "131",
    number = "24",
    pages = "241901",
    year = "2023"
}

@article{Caputa:2024xkp,
    author = "Caputa, Pawel and Kutak, Krzysztof",
    title = "{Krylov complexity and gluon cascades in the high energy limit}",
    eprint = "2404.07657",
    archivePrefix = "arXiv",
    primaryClass = "hep-ph",
    reportNumber = "YITP-24-49, IFJPAN-IV-2024-6",
    doi = "10.1103/PhysRevD.110.085011",
    journal = "Phys. Rev. D",
    volume = "110",
    number = "8",
    pages = "085011",
    year = "2024"
}

@article{Hentschinski:2024gaa,
    author = "Hentschinski, Martin and Kharzeev, Dmitri E. and Kutak, Krzysztof and Tu, Zhoudunming",
    title = "{QCD evolution of entanglement entropy}",
    eprint = "2408.01259",
    archivePrefix = "arXiv",
    primaryClass = "hep-ph",
    doi = "10.1088/1361-6633/ad910b",
    journal = "Rept. Prog. Phys.",
    volume = "87",
    number = "12",
    pages = "120501",
    year = "2024"
}

@article{Hatta:2024lbw,
    author = "Hatta, Yoshitaka and Montgomery, Jake",
    title = "{Maximally entangled gluons for any x}",
    eprint = "2410.16082",
    archivePrefix = "arXiv",
    primaryClass = "hep-ph",
    doi = "10.1103/PhysRevD.111.014024",
    journal = "Phys. Rev. D",
    volume = "111",
    number = "1",
    pages = "014024",
    year = "2025"
}

@article{Ouchen:2025tta,
    author = "Ouchen, Mustapha and Prygarin, Alex",
    title = "{Pomeron evolution, entanglement entropy and Abramovskii-Gribov-Kancheli cutting rules}",
    eprint = "2508.12102",
    archivePrefix = "arXiv",
    primaryClass = "hep-ph",
    month = "8",
    year = "2025"
}

@article{Kutak:2025syp,
    author = "Kutak, Krzysztof and Prasza{\l}owicz, Micha{\l}",
    title = "{Entropy, purity and gluon cascades at high energies with recombinations and transitions to vacuum}",
    eprint = "2508.13781",
    archivePrefix = "arXiv",
    primaryClass = "hep-ph",
    reportNumber = "IFJPAN-IV-2025-15",
    month = "8",
    year = "2025"
}

\end{document}